# 5.5-7.5 MeV Proton generation by a moderate intensity ultra-short laser interaction with $H_2O$ nano-wire targets

A. Zigler[1], T. Palchan[1], N. Bruner[1], E. Schleifer[1], S.Eisenmann[1],

Z. Henis[1], M. Botton[1], S.A. Pikuz[2], Jr. A.Y. Faenov[2], D. Gordon[3], P.Sprangle[3]

(1) Racah Institute of Physics, Hebrew University, Jerusalem 91904, Israel

(2) Joint Institute for High Temperatures, Russian Academy of Sciences, Izhorskaja Street 13/19, 127412, Moscow, Russia

(3) Plasma Physics Division, Naval Research Laboratory, Washington, D.C. 20375, USA

Abstract

*We report on the first generation of 5.5-7.5 MeV protons by a **moderate intensity short-pulse laser** (4.5 × $10^{17}$ W/cm$^2$, 50fsec) interacting with **$H_2O$ nano-wires** (snow) deposited on a Sapphire substrate. In this setup, the laser intensity is locally enhanced by the tip of the snow nano-wire, leading to high spatial gradients. Accordingly, the plasma near the tip is subject to enhanced ponderomotive potential, and confined charge separation is obtained. Electrostatic fields of extremely high intensities are produced over the short scale length, and protons are accelerated to MeV-level energies*.

Compact sources of high energy protons (50-500 MeV) are expected to be key technology in a wide range of scientific applications [1-8]. Particularly promising is the target normal sheath acceleration (TNSA) scheme [9,10], holding record level of 67 MeV protons generated by a peta-Watt laser [11]. In general, laser intensity exceeding $10^{18}$ W/cm$^2$ is required to produce MeV level protons. Enhancing the energy of generated protons using compact laser sources is very attractive task nowadays. Recently [12,13], 1-3 MeV ions were produced by $10^{17}$ -$10^{19}$ W/cm$^2$ lasers illuminating nano-scale targets. In both cases, ion emission was found to be strongly anisotropic. Previous study of snow nano-wires target [14,15] illuminated by moderate energy level short-pulse laser, have demonstrated a significantly improved absorption of the laser energy by the snow deposited target compared to Sapphire only targets.

In this letter, we report for the first time generation of 5.5-7.5 MeV protons by a modest energy laser. The protons are obtained from a nano-wires (snow) target that is positioned at the focal point of a 0.5 TW, 40 fsec laser at the Hebrew University

High Intensity Laser facility (see Fig. 1a). The laser operates at a central wavelength of 798 nm, and is focused by an off-axis parabolic mirror (F# = 3.3) to a spot area of 80 μm² (FWHM) on the target. The laser irradiates the snow interface at 60deg to the normal. A pre-pulse of time duration of the same time scale as the main pulse, originates in the regenerative pre-amplifier, and precedes the main pulse by 10 nsec. The main pulse to pre-pulse contrast ratio is $10^3$. A microscopic imaging system is used to make sure that the laser irradiates a fresh patch of snow at every shot. The diagnostics of the accelerated protons consist of CR39 SSNTD plates, covered by various layers of B10 and Aluminum foils serving as energy filters. The stopping power of each filter setup was calculated using SRIM code [16]. The detector is positioned at a distance of 35 mm from the target's plane along the normal, collecting protons from solid angle of 1.2 sr. A reference CR39 detector is placed at a position hidden from the target to measure the background signal. Each detector with energy filter setup is exposed to a series of 30-50 laser shots before being pulled out of the vacuum chamber and processed.

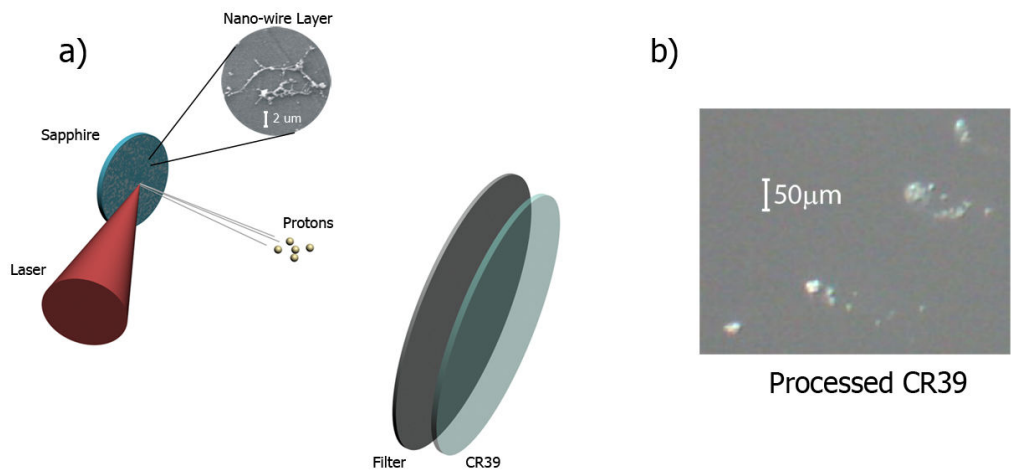

FIG. 1 (color online). (a) Schematics of experimental setup. (b) Proton track bunches on a processed CR39 plate.

A typical scan of the processed CR39 detector is shown in Fig. 1b. The proton tracks (bright, circular spots) are accumulated in bunches, with no preferred orientation on the detector plane. Proton bunches are accelerated into relatively small solid angles of about 1.6 μsr. We therefore deduce that a bunch represents several protons accelerated in one shot at the same direction. Considering the stopping power of the aluminum foil (of thickness 208μm) that is placed in front of the CR39 detector, we conclude that the energy of the fastest protons is 5.5-7.5 MeV. The number of proton tracks behind the filters is 4708 ± 707 protons × shot$^{-1}$ × sr$^{-1}$ compared to a background level of 2942 ± 1079 on the unexposed CR39. Note that the detector is placed at an angle to the laser beam thus the protons are not accelerated along the laser direction as in the planar TNSA experiments. The scaling of the protons energy as a function of the intensity of the main pulse is shown in Fig. 2. The data obtained in the current experiments (HU) follows the conventional relation $E_{max} \propto \left(I\lambda^2\right)^{0.5}$, but is achieved with lower energy and intensity levels of the laser.

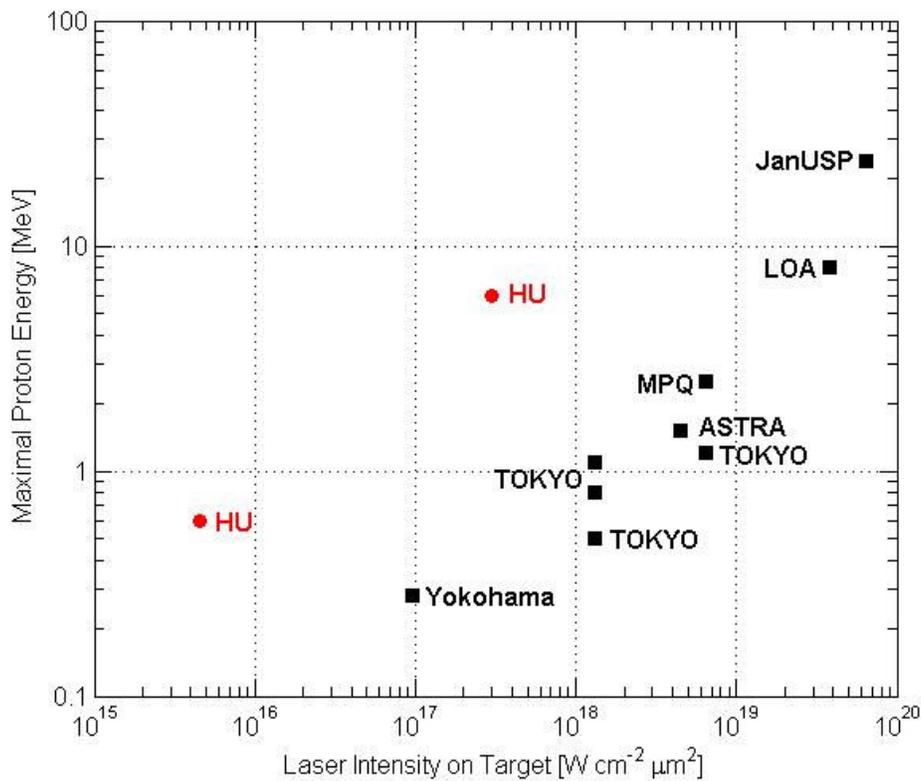

FIG. 2 (color online). Maximal proton energy in MeV versus the intensity for a short <100 fsec Ti: Sapphire laser systems (Data taken from Ref [3,17])

A possible explanation of the experimental results starts with the assumption that the arrangement of the nano-wires on the target is amorphous due to the deposition method. Based on our previous analysis of the snow target [14,15] we estimate that the distance between adjacent nano-wires (~5-10 μm) is large compared to their diameter (~0.1 μm) (see Fig 1a). Furthermore, the spot size of the laser beam is small enough so that it interacts with few nano-wires. For simplicity, we assume that the laser (pre-pulse and main pulse) interacts with only one nano-wire. The highly efficient absorption of the laser energy by the snow target [14] points to the fact that the pre-pulse completely vaporizes the nano-wire (note that the plasma skin depth is larger than the width of the nano-wire). The temperature of the formed plasma is estimated to be about 2-5 eV. During the 10 nsec interval between the pre-pulse and the main pulse, the plasma freely expands away from the nano-wire. The main pulse therefore interacts with a non-uniform cylindrical symmetric plasma density (see Fig. 3), unlike the conventional TNSA foil configuration where the laser interacts with constant or planar plasma distribution.

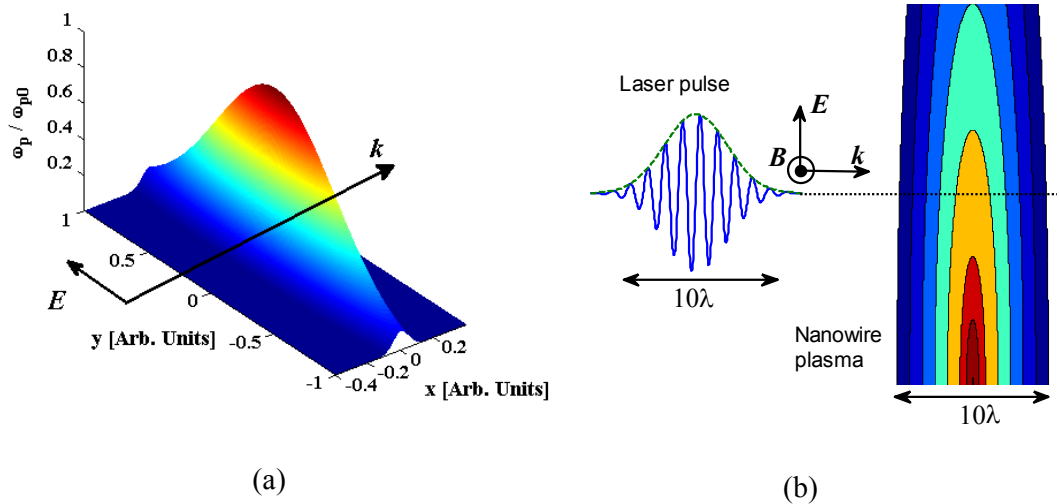

(a)  (b)

FIG. 3 (color online). (a) The plasma density as a function of planar distance from the nano-wire (positioned at x=0, y=0). (b) The laser electric field, **E**, propagation through the nano-wire plasma at a direction **k**.

For brevity of presentation, the main laser pulse is assumed to propagate in a direction perpendicular to the nano-wire symmetry axis. Even with these simplifications the interaction is essentially three dimensional as the laser breaks the cylindrical symmetry of the nano-wire. The main-pulse length is larger than the width of the nano-wire. Accordingly the averaged ponderomotive potential of the laser in the vicinity of the tip can be calculated using the quasi-static approximation. We model the tip as a prolate spheroid with conductivity and dielectric constant related to the plasma parameters. The main result of this model is the field enhancement near the tip by an amount related to the tip radius and the distance from the tip. To obtain a clear, yet representing, one dimensional model of the acceleration process, we analyze the interaction of the laser with the plasma distributed along one ray in parallel to the laser wave-vector. The density profile of the plasma along a ray and the enhancement factor of the laser amplitude due to the tip are described in Fig. 4. We now solve a set of one-dimensional fluid equations for the electrons and ions in a plasma of the given distribution function assuming that the main pulse propagates from the left edge to the right and is subject to the tip enhancement. The effect of the ponderomotive potential in this case is enhanced by two factors. First is the local field enhancement which is translated to direct enhancement of the amplitude. Next is the increment of the gradient of the ponderomotive potential due to the local character of the field enhancement. The plasma electrons are therefore subjected to a greatly increased force and accordingly the density is modified to a greater extent than is expected of the laser intensity by itself. Fig. 5 demonstrates this enhancement for six time frames. Solid lines shows the electrons density at the given time, broken lines are the initial density of the electrons (namely the ions density), and dotted lines represent the laser envelope. Note that as the laser pulse is near the tip, its shape is strongly affected by the local field enhancement (Fig. 5c,d). The electrons respond to the ponderomotive potential and the self consistent electrostatic potential, and their density is changed as the pulse propagates through the plasma.

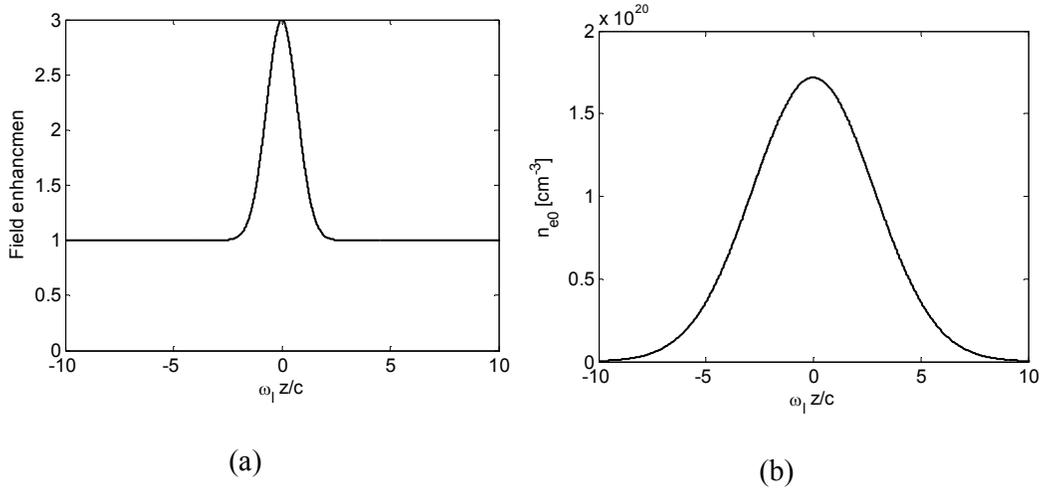

FIG. 4. (a) The localized field enhancement due to the tip. (b) The initial plasma density.

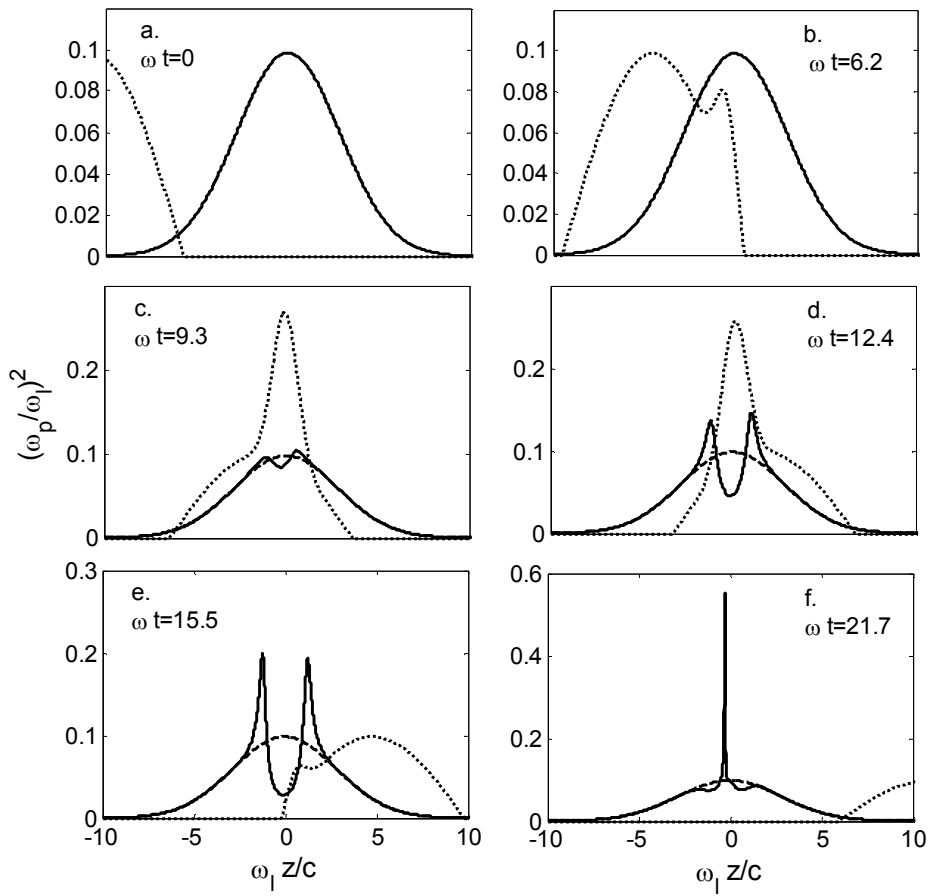

FIG. 5. The density of the electrons (solid line) at six time frames. The laser envelope (dotted line) is enhanced by the tip. Broken line shows the initial distribution of the electrons.

Fig. 6 shows the distribution of the electrons (black line) after the passage of the laser pulse including the tip enhancement. The exact position and magnitude of the electron cloud is a function of the maximal enhancement and gradient of the tip. We note that the level to which the electrons are compressed depends also on the temperature they gain in the process. Up to this stage the ions are practically not affected by the laser pulse at all. However, they start to respond to the electrostatic potential of the electrons cloud. Following the conventional model [18], we estimate the temperature of the electrons in the cloud, the scale length that determines the intensity of the accelerating field and the accelerating distance. To estimate the temperature of the electrons we use the following expression:

$$kT_{Hot} \sim m_e c^2 \left[ \sqrt{1 + \frac{a_{tip}^2 I \lambda^2}{1.37 \times 10^{18}}} - 1 \right]$$

Here the conventional scaling of the laser intensity is multiplied by the field enhancement of the tip, $a_{tip}$ (squared for intensity). Taking this factor to be 10 (which means a modest field enhancement of ~3) we find that the temperature of the hot electrons in the cloud is of the order of 200-300 keV. Next we estimate the scale length of the local plasma using the calculations of the cold fluid model. Based on the fluid model we find that the electrons cloud is ramped up by a steep gradient which is estimated to be 0.1-0.05 λ (see left inset in Fig. 6). Furthermore, we estimate that the length over which the protons can be accelerated is approximately λ (see right inset in Fig. 6). Combining the above we find that the protons can be accelerated to about 10-20 $kT_{hot}$ which for laser intensity at the range of 4.5 × 10$^{17}$ W/cm$^2$ is 2-6 MeV. Several remarks are in order. First, we would like to stress that the most energetic protons are accelerated in a direction that is defined by the orientation between the tip and the wave vector of the laser. This is supported by our experimental results where we find bunches of accelerated protons and not a uniform distribution. Furthermore, as this acceleration scheme is ballistic in nature we do not expect a large amount of protons to reach this energy unless specially designed targets are used. Finally, three-dimensional effects, which will not change the basics of the accelerating scheme, must be considered to describe the exact position of the electron cloud with respect to the tip depending on the parameters of the plasma, the tip and the laser.

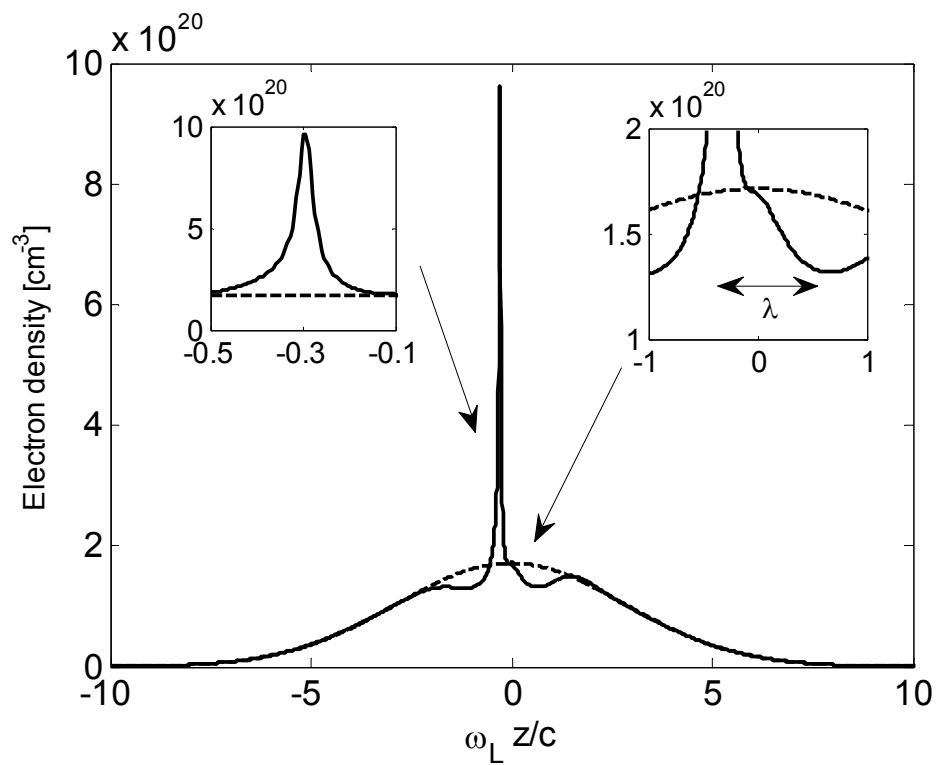

FIG. 6. Electron density normalized to the laser frequency before and after the main laser pulse has passed the $H_2O$ nano-wire. Left inset: Zoom on the region of peak electron density. Right inset: Zoom on the ions acceleration length.

In conclusion, we have experimentally demonstrated generation of 5.5-7.5 MeV protons from snow nano-wire targets by modest, $4.5 \times 10^{17}$ W/cm$^2$ laser intensities, which is an order of magnitude higher compared with previous experiments with such laser intensities. The protons are accelerated by the enhanced interaction of the laser field and the plasma near the tip of the nano-wire. Engineered nano-wire targets can improve the interaction scheme by increasing the number of accelerated protons and directing them in a pre-designed direction.